\documentclass[structabstract]{aa}  
\usepackage{graphicx}
\usepackage{txfonts}
\usepackage{natbib}
\usepackage{color}

\def\spose#1{\hbox to 0pt{#1\hss}}
\def\gsim{\mathrel{\spose{\lower 3pt\hbox{$\mathchar"218$}}
          \raise 2.0pt\hbox{$\mathchar"13E$}}}
\def\lsim{\mathrel{\spose{\lower 3pt\hbox{$\mathchar"218$}}
          \raise 2.0pt\hbox{$\mathchar"13C$}}}

\begin{document}

\title{The $E_{\rm p}$ - $E_{\rm iso}$ relation and the internal shock model}

\author{R. Mochkovitch$^1$ \and  L. Nava$^{2,3}$}

\authorrunning{R. Mochkovitch \and  L. Nava}
\titlerunning{The $E_{\rm p}$ - $E_{\rm iso}$ relation and the internal shock model}

\institute{$^{1}$ UPMC-CNRS, UMR7095, Institut d'Astrophysique de Paris, F-75014, Paris, France\\
$^{2}$ Racah Institute of Physics, The Hebrew University of Jerusalem, 91904, Israel\\
$^{3}$ APC, Universit\'e Paris Diderot, CNRS/IN2P3, CEA/Irfu, Observatoire de Paris, Sorbonne Paris Cit\'e, France}

\abstract
{The validity of the $E_{\rm p}$ - $E_{\rm iso}$ correlation in gamma-ray bursts and the possibility of explaining the prompt emission
with internal shocks are highly debated questions. }         
{We study whether the $E_{\rm p}$ - $E_{\rm iso}$ correlation can be reproduced if internal shocks are indeed responsible for the prompt emission, or
conversely, if the correlation can be used to constrain the internal shock scenario.}
{We developed a toy model where internal shocks are limited to the collision of only two shells. Synthetic burst populations
were constructed for various distributions of the
model parameters, such as the injected power in the relativistic outflow, the average Lorentz factor, and its typical contrast between 
the shells. These parameters can be independent or linked by various relations.}
{Synthetic $E_{\rm p}$ - $E_{\rm iso}$ diagrams are obtained in the different cases and compared with the observed correlation. 
The reference observed correlation is the one defined by the BAT6 sample, a sample of Swift bursts almost complete
in redshift and affected by well-known and reproducible instrumental selection effects. 
The comparison is then performed with a subsample of synthetic bursts that satisfy the same selection criteria as were imposed on the BAT6 sample.
A satisfactory agreement between model and data can often be achieved, but only if several strong constraints are satisfied on both the dynamics 
of the flow and the microphysics that governs the redistribution of the shock-dissipated energy. }
{}

 \maketitle

\section{Introduction}
The origin of the prompt emission of gamma-ray bursts (hereafter GRBs) is still debated. Temporal variability down to very
short time-scales imposes that the emission comes directly from the relativistic outflow emitted by the central engine
and not from its interaction with the circumburst medium {\citep[external origin;][]{SP97}}. But at least three possibilities remain 
for an internal origin: {\it (i)} dissipation below the photosphere that can modify the emerging thermal spectrum by inverse
Compton scattering off energetic electrons \citep{RM05,PMR05,PEER2011,GIANNIOS2012,BELO2013};
{\it (ii)} dissipation above the photosphere either of the flow kinetic energy by internal shocks \citep{RM94,KPS97,DM98}, or {\it (iii)} 
in a magnetized ejecta through 
reconnection processes \citep{McKU2012,YZ2012,ZZ2014}. Many predictions on the light curves and spectra of GRBs can be made from internal shocks
\citep{BD14}. Several agree well with observations, but the shape of the expected synchrotron spectrum does not 
fit well, because it is too soft at low energy (\citealt{PREECE98,GCL00}; see however \citealt{DERISHEV07,DBD11}). Moreover, 
the necessary efficient transfer 
of dissipated energy to electrons has been disputed for a moderate 
magnetization of the flow $\sigma>0.1$, where $\sigma$ is the ratio of the Poynting flux to the particle rest energy flux \citep{MA2010,NKT11}. 

Photospheric dissipation and reconnection models have been proposed to avoid these problems. Photospheric dissipation 
can take place through radiation mediated shocks \citep{Levin2012,KL2014}, collisional heating \citep{BELO2010}, or reconnection, and then
the main emission mechanism is not synchrotron. 

If reconnection takes place above the photosphere, the emission should again come from the 
synchrotron process, as it does for internal shocks with the same potential problems regarding the spectral shape. For photospheric 
and reconnection models few works have been dedicated to actually 
produce light curves that can be compared with data to test the temporal (and spectro-temporal) evolution of the models 
(see \citealt{ZZ2014}, however).      

Scenarios for the prompt emission can be tested on their ability to reproduce not only light curves and spectra of individual 
events, but also the properties of the GRB population as a whole. An example is the $E_{\rm p}$ - $E_{\rm iso}$ (or Amati) relation \citep{AMATI02}
or $E_{\rm p}$ - $E_{\gamma}$ (or Ghirlanda) relation \citep{GHIRL04}, where $E_{\gamma}$ is the true energy release in gamma-rays. 

Similarly to the possibility of explaining the prompt emission of GRBs with internal shocks, the validity of  
the Amati relation has been disputed with indications that it might be, at least partially,
the result of selection effects \citep{NP2005,BP2005,BUTLER07,SN2011}.
However, several studies, aimed at quantifying these selection effects for different detectors, have shown that even if
instrumental biases contribute to shaping the distribution of GRBs in the $E_{\rm p}$ - $E_{\rm iso}$ plane, they cannot be fully 
responsible for the observed correlation, which therefore must also have a physical origin \citep{GHIRL08,NAVA08,GHIRL12}.   

Attempts have been made 
to interpret the Ghirlanda relation as the result of viewing-angle effects \citep{LE05}
or of similar comoving-frame properties for all GRBs, with correlations
being the results of the spread in the jet Lorentz factor \citep{GNGC12}.
The estimate of the collimation-corrected energy $E_\gamma$ requires measuring
the jet break-time from late-time afterglow observations and the knowledge of the
density of the circum-burst medium. Depending on the assumed density profile, the
slope of the $E_{\rm p}-E_\gamma$ correlation is around 0.7 for a homogeneous
density profile and around 1 for a wind-like density profile (Nava et al., 2006). In
both cases the slope is steeper than the slope of the Amati correlation. The
consistency between the $E_{\rm p}-E_\gamma$ and $E_{\rm p}-E_{\rm iso}$
correlations and their different slopes and scatters can be explained by assuming that
the jet opening angle is anticorrelated with the energy \citep{GHIRL05,GHIRL13}. 

In this paper we focus on the $E_{\rm p}-E_{\rm iso}$ correlation.
We especially wish to determine {\it (i)} under which conditions the internal shock model would be able 
to account for it, and {\it (ii)} to exa\-mine whether these conditions are realistic and 
can indeed be satisfied.
To do this we also take into account the role of selection effects in the observed $E_{\rm p}$ - $E_{\rm iso}$ relation. 
While the lack of bursts with a high $E_{\rm iso}$ and a low $E_{\rm p}$ should have a physical origin, events with a low $E_{\rm iso}$ 
and a high $E_{\rm p}$ may escape detection, as discussed by \cite{HA13}.

The paper is organized as follows: we describe in Sect.2 the toy model we used to generate large populations of synthetic bursts
with different assumptions on the model parameters and the possible links between them. We discuss in Sect.3 the role of instrumental biases
with the aim to construct a synthetic sample including selection effects similar to those affecting a reference sample of observed bursts. 
We then compare the $E_{\rm p}$ - $E_{\rm iso}$ relations defined by various synthetic populations with the relation defined by the reference sample. 
Our results are discussed in Sect.4, which is also the conclusion.    

\section{Constructing a large population of synthetic bursts}
\subsection{Two-shell internal shock toy model}
To generate a large number (up to $10^6$) of synthetic bursts, we restricted the internal shock phase to the collision of only
two shells. Obviously, using this simplified approach we loose most of the details of the burst temporal evolution, but
we expect to preserve the main features of the energetics and the peak of the time-integrated spectrum, 
which we need to obtain the $E_{\rm p}$ - $E_{\rm iso}$ relation. This 
model has previously been presented in \cite{BARRAUD05}, and we summarize their main assumptions and equations here. 

The two shells have respective masses and Lorentz factors ($M_i$, $\Gamma_i$ with $i=1,2$) and are produced over a total
duration $\tau$. We can then define the average power injected in the relativistic outflow
\begin{equation}
\dot E={(M_1\Gamma_1 + M_2 \Gamma_2)c^2\over  \tau}=\dot M {\bar \Gamma} c^2\ ,
\end{equation}   
where $\dot M=(M_1+M_2)/\tau$ and ${\bar \Gamma}=(M_1\Gamma_1+M_2\Gamma_2)/(M_1+M_2)$ are the average mass loss rate and Lorentz factor.
The collision radius\footnote{We checked that the shock radius is located above the photosphere in all but a few $10^{-3}$ of the
bursts in our Monte Carlo approach of Sect. 2.2.} is
\begin{equation}
R_{\rm s}=2 c t_{\rm v}\,{\Gamma_1^2\Gamma_2^2\over \Gamma_2^2-\Gamma_1^2}\ , 
\end{equation}
where $\Gamma_2 > \Gamma_1$ has been assumed and $t_{\rm v}\le \tau$ is a typical va\-ria\-bility time scale over which 
the bulk of the energy is released, which is of about one second in long bursts \citep{NP02}.  
It is a key parameter
in the internal shock model because it fixes the location of the shocks, and introducing it 
allows us to go somewhat beyond the basic two-shell model. 
   
A fraction $\epsilon_e$ of the dissipated energy is transferred to electrons and radiated so that
\begin{equation}
E_{\rm iso}=\epsilon_e\,E_{\rm diss}=\epsilon_e\,\left[M_1 \Gamma_1+M_2 \Gamma_2-(M_1+M_2)\,\Gamma_{\rm f}\right]c^2\ ,
\end{equation}
where the final Lorentz factor after the two shells have merged is given by
\begin{equation}
\Gamma_{\rm f}=\sqrt{\Gamma_1 \Gamma_2 {M_1\Gamma_1+M_2\Gamma_2\over M_2\Gamma_1+M_1\Gamma_2}}\ .
\end{equation}
The peak energy of the synchrotron spectrum is 
\begin{equation}
E_{\rm p}\sim E_{\rm syn}=C_{\rm syn}\,\Gamma_{\rm f}B\Gamma_e^2\ ,
\end{equation}
where $B$ and $\Gamma_e$ are the post-shock magnetic field and electron Lorentz factor and 
$C_{\rm syn}={3\over 4\pi}{e h\over m_e c}$. We obtain $B$ and $\Gamma_e$ using
the redistribution parameters $\epsilon_e$, $\epsilon_B$ and $\zeta$ (fraction of electrons that are accelerated)
\begin{equation}
B\sim (8\pi \epsilon_B\,\rho e )^{1/2}\ \ \ {\rm and}\ \ \ \Gamma_e\sim {\epsilon_e\over \zeta}{m_p\over m_e}{e\over c^2}\ ,
\end{equation}
where
\begin{equation}
\rho\sim {\dot M\over 4\pi R_{\rm s}^2 \,{\bar \Gamma} c}={{\dot E}\over 4\pi R_{\rm s}^2 \,{\bar \Gamma}^2 c^3}
\end{equation}
is the post-shock density and 
\begin{equation}
e={E_{\rm diss}\over (M_1+M_2)\Gamma_{\rm f}}
\end{equation} 
is the dissipated energy per unit mass in the comoving frame.
Equations (7) and (8) lead to  
\begin{equation}
E_{\rm p}=C_{\rm p}\,\Gamma_{\rm f}\,\rho^{1/2} e^{5/2}
\end{equation}
with
\begin{equation}
C_{\rm p}=C_{\rm syn}\,(8\pi \epsilon_B)^{1/2}\left({\epsilon_e\over \zeta c^2}{m_p\over m_e}\right)^2\ .
\end{equation}
If we additionally assume for simplicity that $M_1=M_2$, the isotropic energy of the burst
$E_{\rm iso}$, its average luminosity $\langle L_{\rm iso}\rangle =E_{\rm iso}/\tau$, and $E_{\rm p}$ can be simply expressed in terms of the model parameters. 
We have  
\begin{equation}
E_{\rm iso}=\epsilon_e\,{\dot E} \tau\,f(\kappa)\ \ \ {\rm ,}\ \ \ \langle L_{\rm iso}\rangle=\epsilon_e\,{\dot E}\,f(\kappa)
\end{equation}
and
\begin{equation}
E_{\rm p}\propto {{\dot E}^{1/2}\,\varphi(\kappa)\over t_{\rm v}\ {\bar \Gamma}^2}\ ,
\end{equation}
where $\kappa=\Gamma_2/\Gamma_1$ and $f$ and $\varphi$ are functions of $\kappa$ only
\begin{equation}
\left\lbrace\begin{array}{cl}
& f(\kappa)={(\sqrt{\kappa}-1)^2\over 1+\kappa}\\
& \varphi(\kappa)={\left[\left(\kappa^2-1\right)\,\left(1+1/\kappa\right)^2\right]\left(\kappa^{1/2}
+\kappa^{-1/2}-2\right)^{5/2}\over \kappa^{1/2}+\kappa^{-1/2}}\ .\\
\end{array}\right.
\end{equation}
\subsection{Monte Carlo approach}
To generate an $E_{\rm p}$ - $E_{\rm iso}$ diagram that can be compared with observations we need to fix the following:
\begin{itemize}
\item the distribution in redshift of the events: we adopted a burst rate that follows the star formation rate SFR3 
of \cite{PM2001}, which increases 
at large $z$, in contrast to the cosmic SFR, which is probably declining. This accounts for the fact that the 
stellar population at large $z$ appears to be more efficient in producing GRBs than at low $z$, so that the burst rate is 
not directly proportional to the SFR \citep{RDM06,KIS09,WP10,BUTLER10}.
\item the distribution of intrinsic duration $\tau$: we adopted a log-normal distribution centered at $\tau=8$ s and 
checked a posteriori that the distribution of the duration of the detected bursts (including
time dilation) agrees with the observed distribution for long GRBs \citep{BATSECAT99}. Similarly, we also adopted a log-normal distribution for the
variability time-scale $t_{\rm v}$, so that the distribution of $t_{\rm v}$ in detected bursts fits that of pulse widths \citep{NP02}.   
\item the distribution of injected power ${\dot E}$: we adopted a power law of index $\delta=-1.6$,
the value of $\delta$ being constrained in the interval $-1.7<\delta<-1.5$ to reproduce the ${\rm Log}\,N$ - ${\rm Log}\,P$ curve
\citep{RDM06}.
The upper limit ${\dot E}_{\rm max}$ must be high enough to make the most energetic events that can exceed 
$E_{\rm iso}=10^{54}$ erg, and we therefore took ${\dot E}_{\rm max}=3\,10^{54}$ erg.s$^{-1}$ to account for the 
low efficiency of internal shocks 
\begin{equation}
{E_{\rm iso}\over \dot E\,\tau}=\epsilon_e\times f(\kappa)\lsim 0.1\ .
\end{equation} 
The lower limit ${\dot E}_{\rm min}$ could be more than six orders of ma\-gni\-tude lower in bursts such as GRB 980425 and
GRB 060218, but these events probably belong to a different po\-pu\-la\-tion
with its own distinct luminosity function \citep{VLZ09}. For cosmological
bursts ${\dot E}_{\rm min}$ is weakly cons\-trained by
observations. We adopted 
${\dot E}_{\rm min}=10^{52}$ erg.s$^{-1}$.       
\item the distributions of contrast $\kappa=\Gamma_2/\Gamma_1$ and a\-ve\-rage Lorentz factor ${\bar \Gamma}$:
the function $\varphi(\kappa)$ in Eq.(13) rapidly increases with $\kappa$ (approximately as $\kappa^5$ for $\kappa\sim 5$)
so that to avoid a too high dispersion
in the $E_{\rm p}$ - $E_{\rm iso}$ relation, $\kappa$ has to be confined within a relatively narrow interval. Unless 
otherwise stated, we adopted a normal distribution for $\kappa$, centered at $\kappa=5$ with a standard deviation
$\sigma_{\kappa}=1$. For ${\bar \Gamma}$, we 
first assumed a uniform distribution from 100 to 400. 
\end{itemize}
These choices correspond to a situation where the various model parameters are not correlated, but we also  
consi\-de\-red the possibility that some of them are directly linked. From studies of the rise time of
the optical afterglow light curve it has been suggested for example, that the average
Lorentz factor increases with burst luminosity as ${\bar \Gamma}\propto L^{1/2}$ \citep{LIANG2010,GNGC12,LU2012}. In this work,
we replaced the luminosity by the injected power and tested the relation
\begin{equation}    
 \bar \Gamma\propto {\dot E}^{1/2}\ .
\end{equation}
Other examples may consist to link the time scale $t_{\rm v}$ or/and amplitude of the fluctuations of the Lorentz factor $\kappa$,
to the a\-ve\-rage Lorentz factor $\bar \Gamma$, that is to assume that the flow becomes more chaotic when it is more relativistic.   
A first possibility, suggested by Eq.(12), would be to have
\begin{equation} 
t_{\rm v}\propto {\bar \Gamma}^{-2}\ ,
\end{equation}
so that $E_{\rm p}\propto {\dot E}^{1/2}\,\varphi(\kappa)$, directly yielding an Amati-like relation
if $\varphi(\kappa)$ does not vary too much.
If additionally Eqs.(15) and (16) are satisfied together, variability and luminosity become connected with
\begin{equation} 
t_{\rm v}\propto {\dot E}^{-1}\ ,
\end{equation}
implying that more luminous bursts will be both more relativistic and more variable \citep{FENIMORE01}.
Similarly, for the contrast in Lorentz factor we tested relations of the form
\begin{equation}
\kappa\propto {\bar \Gamma}^{\nu}\ .
\end{equation}    

To compute $E_{\rm p}$ and $E_{\rm iso}$, we finally fixed the values of the microphysics parameters: we took
$\epsilon_e=0.3$, $\epsilon_B=0.01$ and $\zeta=3\,10^{-3}$ (i.e., $\epsilon_e/\zeta=100$). The low value of 
$\zeta$ is required to guarantee that the emission occurs in the soft gamma-ray range.

\section{Producing a synthetic $E_{\rm p}$ - $E_{\rm iso}$ diagram}
\subsection{Selection effects}
After $\dot E$, $\tau$, $\bar \Gamma$, $\kappa$ and the redshift were drawn according to the assumed distributions,
we know for each synthetic burst $E_{\rm iso}$ (Eq.11) and $E_{\rm p}$ (Eq.12).
Adopting a Band shape, we then computed the average flux and the fluence received on Earth in any spectral interval. 
The low and high-energy spectral indices were fixed to $\alpha=-1$ and $\beta=-2.5$, which corresponds to the mean values of 
the observed distributions \citep{PREECE00}. Synchrotron emission in
the fast-cooling regime instead predicts $\alpha=-1.5$, but including the inverse-Compton process and a decreasing magnetic field
behind the shocks can help to reduce the discrepancy \citep{DERISHEV07,DBD11}. 
Adopting $\alpha=-1.5$ or $-1$ for the present study leads to very similar results. 

To compare synthetic and observed $E_{\rm p}$ - $E_{\rm iso}$ sequences, we have to apply to the synthetic sample the very same selection effects
that affect the observed sample. The main selection effects arise from the requirement
to trigger the event, measure $E_{\rm iso}$, $E_{\rm p}$ (which must fall inside the range of sensitivity of the instrument),
and the redshift. The trigger threshold
can be approximated as a threshold on the peak flux, while the need to perform a
good spectral analysis broadly translates into a limit on the fluence (spectral
threshold). \cite{GHIRL08}, \cite{NAVA08}, and \cite{NAVA11} discussed these effects
in detail, and derived for each of the relevant instruments the trigger
and spectral thresholds. To lie above the spectral threshold is typically a more demanding request
than to lie above the trigger threshold: to detect a burst is not a sufficient
condition to derive $E_{\rm p}$ and $E_{\rm iso}$ from the spectral
analysis.

To understand whether internal shocks can reproduce the observed correlation, we 
introduced in the sample of synthetic bursts the selection
effects that affect the sample of observed bursts.
This is a hard task, given the complexity of selection effects and the fact that
the observed bursts have been detected by different instruments, which introduce
different thresholds.
To study this properly, we compared our population of synthetic
events with a sample with well-known and reproducible instrumental selection effects.
We chose the BAT6 sample defined by Salvaterra et al. (2012).
This is a subsample of the GRBs detected by BAT, which includes events with
favorable observing conditions and with a peak flux $F_{\rm peak}> 2.6$ ph.cm$^{-2}$s$^{-1}$ 
in the 15 - 150 keV energy range. These requirements resulted in a sample of 58
GRBs with a redshift-completeness level of almost 90\% \citep{SALV12}\footnote{The redshift completeness of the BAT6 sample has now been increased
to 95\% \citep{COV13}.}.
The value of the flux threshold
was chosen to reach a good compromise between redshift completeness
and number of events, that is, still high enough to perform statistical studies.
However, this value has another main advantage: it ensures that all the events
detected above this threshold (which is much higher than the BAT trigger threshold)
also lie above the spectral threshold. This means that for all bursts above that
flux, the spectral analysis can be performed and the spectral threshold does not
introduce strong effects. The dominant selection effect is the flux threshold, which
for this sample is well known. For an appropriate comparison, we just
applied the very same selection criteria to our sample of synthetic bursts.

From the whole sample of simulated bursts, we then selected only those with $F_{\rm peak}> 2.6$ 
ph.cm$^{-2}$.s$^{-1}$  in the 15 - 150 keV energy range and compared their properties
in the $E_{\rm p}-E_{\rm iso}$ plane with the 50 events of the BAT6 sample with a measured redshift
{\citep[the properties of the BAT6 sample in the $E_{\rm iso}$ - $E_{\rm p}$ plane have been studied in][]{NAVA12}}.
Our simple model only provides the average flux $\langle F\rangle$ of each synthetic burst however, which can be much lower 
than the peak flux in a highly variable event. 
To obtain an estimate of $F_{\rm peak}$ we then applied a correction factor to $\langle F\rangle$.
From inspecting long GRBs detected by BATSE, we found that 
in the plane $F_{\rm peak}/{\langle F\rangle}$ vs. $T_{90}$, these GRBs are distributed inside a triangular
region. The lower and upper edges of this region are approximately described by the relations ${F_{\rm peak}/ {\langle F\rangle}}=T_{90}^{0.2}$
and $F_{\rm peak}/ {\langle F\rangle}=T_{90}^{0.6}$, so that, to convert average fluxes into peak fluxes, we used the expression 
\begin{equation}
{F_{\rm peak}\over {\langle F\rangle}}=\left[(1+z)\tau\right]^{[0.4+0.4(R-0.5)]}\ ,
\end{equation}  
where the random variable $R$ is uniformly distributed between 0 and 1. 
\subsection{Results}
The resulting $E_{\rm p}$ - $E_{\rm iso}$ synthetic sequences are shown in Fig.1 together with the observed BAT6 sample
in the following cases (except {\it ii}):
\begin{itemize}
\item({\it i}) no correlation between model parameters: a power-law fit of the resulting sequence yields $E_{\rm p}=136\,E_{\rm iso,52}^{0.57}$ keV 
with a dispersion of 0.4 in the ${\rm Log}\,E_{\rm iso}$ - ${\rm Log}\,E_{\rm p}$ plane.  
\item ({\it ii}) $\bar \Gamma= 40\,{\dot E}_{52}^{1/2}$ with a dispersion of 0.3 in ${\rm Log}\,\bar \Gamma$: the predicted correlation 
is opposite to the observed correlation with $E_{\rm p}\propto E_{\rm iso}^{-1.1}$. 
\item ({\it iii}) $\bar \Gamma= 40\,{\dot E}_{52}^{1/2}$ is now only the lower value of $\bar \Gamma$ for a given $\dot E$, the maximum
being ${\bar \Gamma}_{\rm max}=700$ and $\bar \Gamma$ is uniformly distributed between these two limits. This agrees with the results of
\cite{HASC14}, who recently recon\-si\-de\-red the estimates of $\bar \Gamma$ from early optical afterglow observations.
It leads to $E_{\rm p}=94\,E_{\rm iso,52}^{0.46}$ keV with a dispersion of 0.38. The resulting sequence is somewhat below the observed sequence.
This can be corrected by reducing the fraction $\zeta$ of accelerated electrons even more. With $\zeta=10^{-3}$ we derive
$E_{\rm p}=147\,E_{\rm iso,52}^{0.51}$ keV with a dispersion of 0.4, which is the sequence represented in Fig.1.
\item ({\it iv}) $t_{\rm v}\propto {\bar\Gamma}^{-2}$; more precisely and to avoid having $t_{\rm v}>\tau$,
we adopted $t_{\rm v}={\rm min}\,[\tau,({\bar \Gamma}/200)^{-2}\ {\rm s}]$
with a dispersion of $0.3$ in ${\rm Log}\,t_{\rm v}$. 
This gives $E_{\rm p}=136\,E_{\rm iso,52}^{0.55}$ keV with a dispersion of 0.36.             
If, in addition to the condition on $t_{\rm v}$, we add the conditions on $\bar \Gamma$ ($\bar \Gamma= 40\,{\dot E}_{52}^{1/2}$ or
$40\,{\dot E_{52}}^{1/2}<{\bar \Gamma}<750$), we obtain very similar results.
\item ({\it v}) $\kappa \propto {\bar \Gamma}^{\nu}$; we illustrate in Fig.1 the choice $\kappa=3\,{\bar \Gamma}_2^{0.5}$ 
(with ${\bar \Gamma}_2={\bar \Gamma}/100$)
and a dispersion of $0.1$ in ${\rm Log}\,\kappa$. We derive $E_{\rm p}=157\,E_{\rm iso,52}^{0.56}$ keV with a dispersion of 0.32.  
\end{itemize}
In all these cases we also obtained the lower limit on the Lorentz factor from the annihilation of photons 
{\citep[correspon\-ding to limit A of][]{LS2001}}. The optically thick bursts are represented in cyan in Fig.1. 
\begin{figure*}
\begin{center}
\begin{tabular}{ccc}
\includegraphics[width=0.5\textwidth]{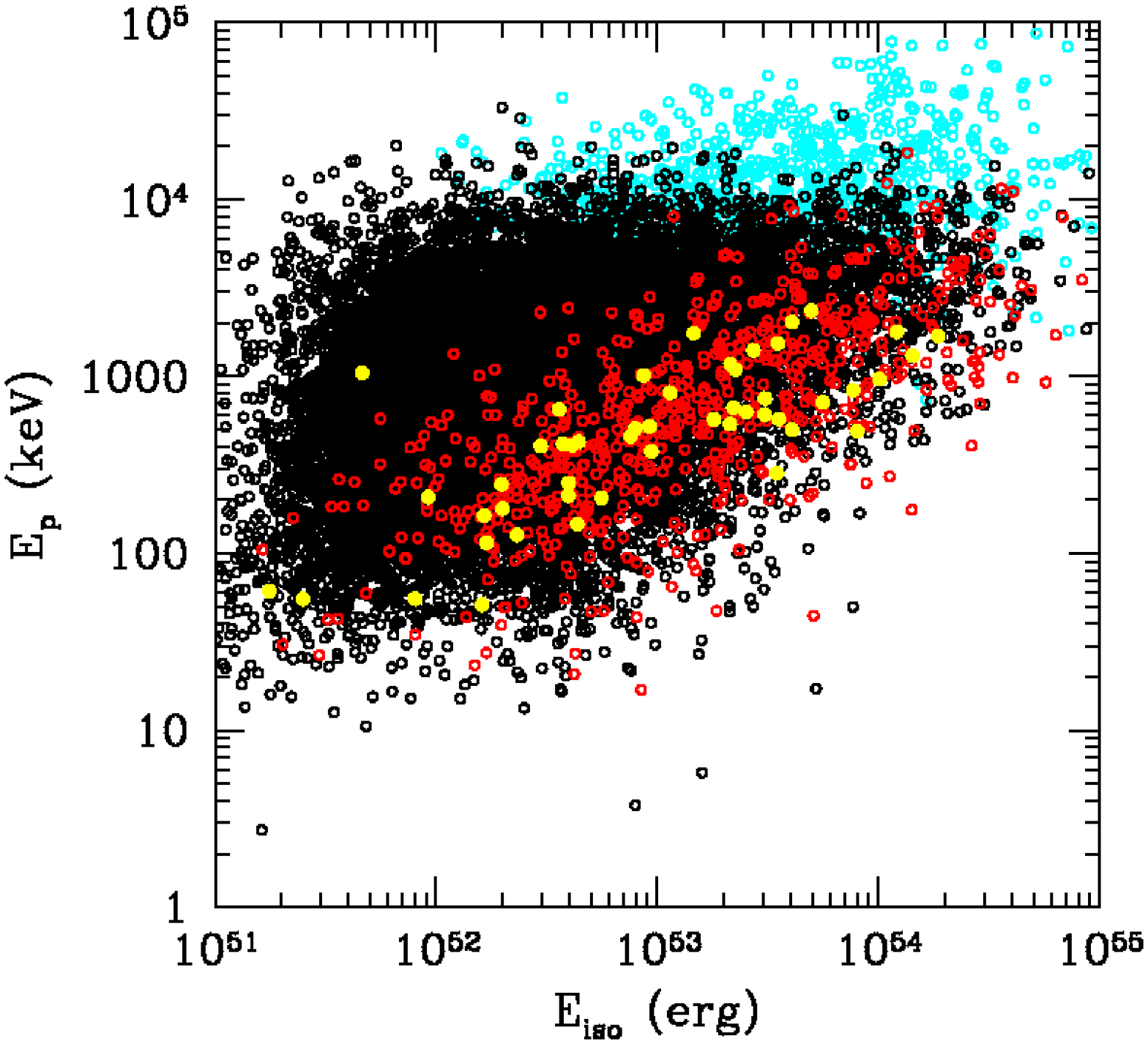} &
\includegraphics[width=0.5\textwidth]{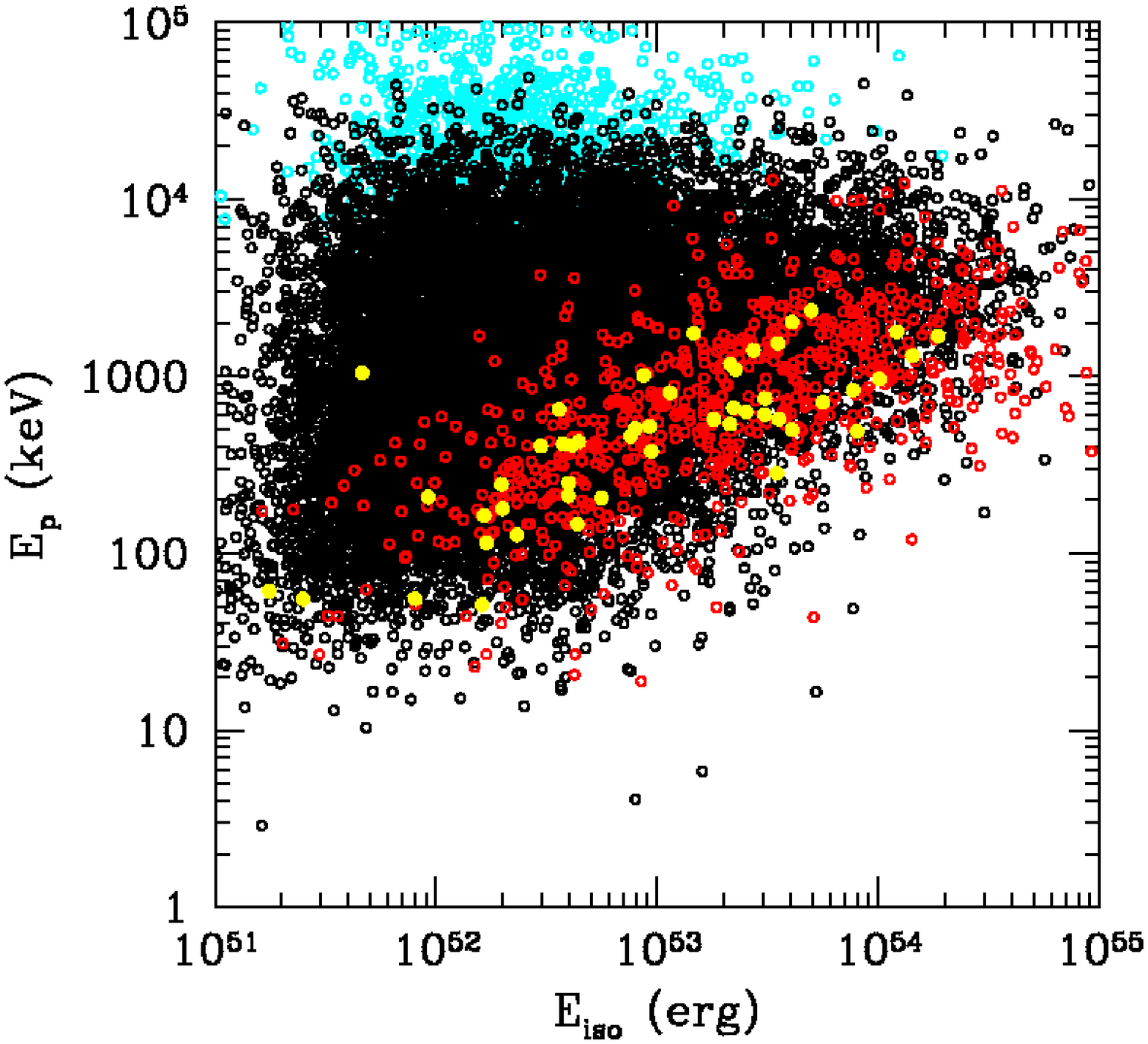} & 
\end{tabular}
\end{center}
\begin{center}
\begin{tabular}{ccc}
\includegraphics[width=0.5\textwidth]{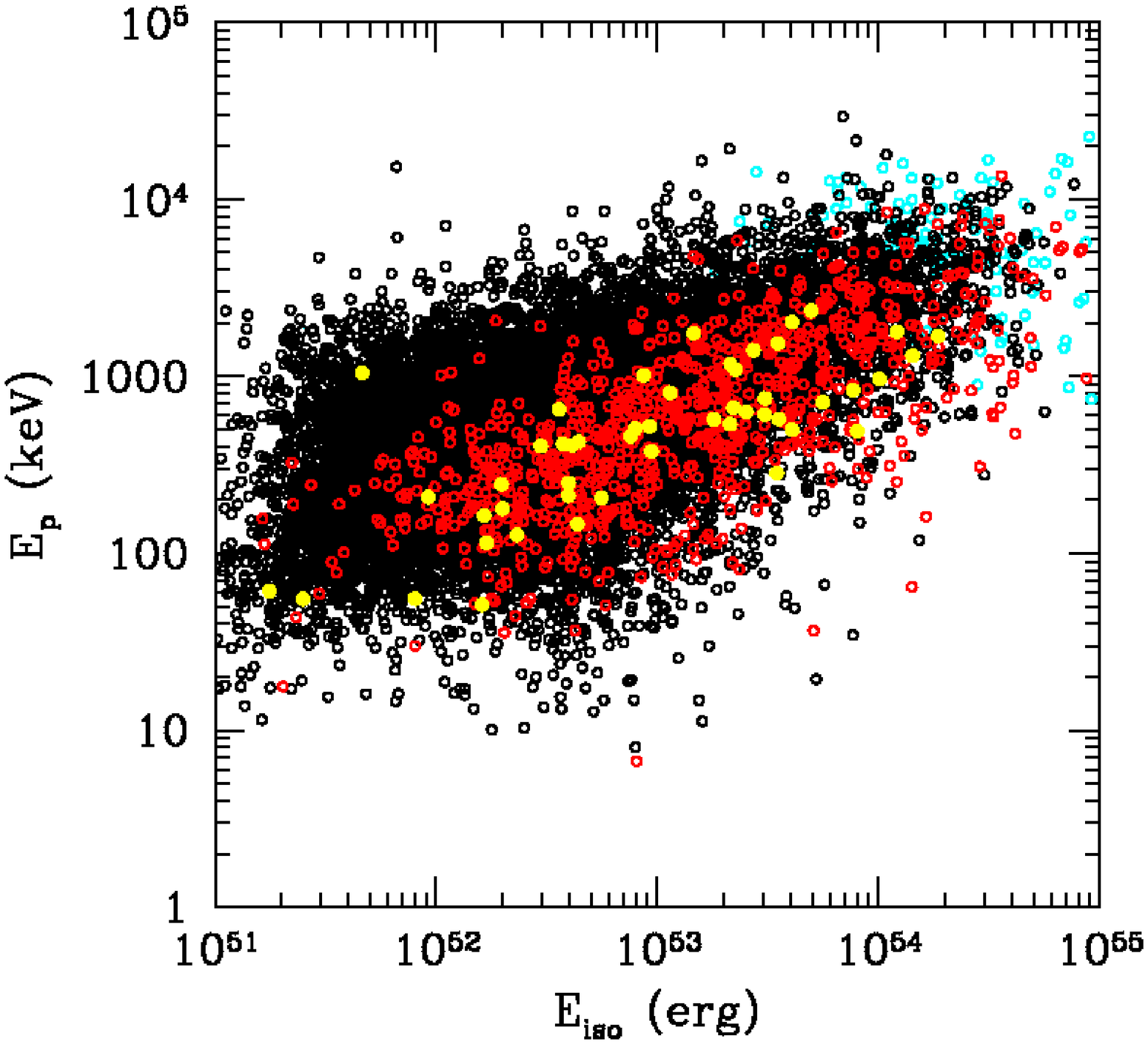} &
\includegraphics[width=0.5\textwidth]{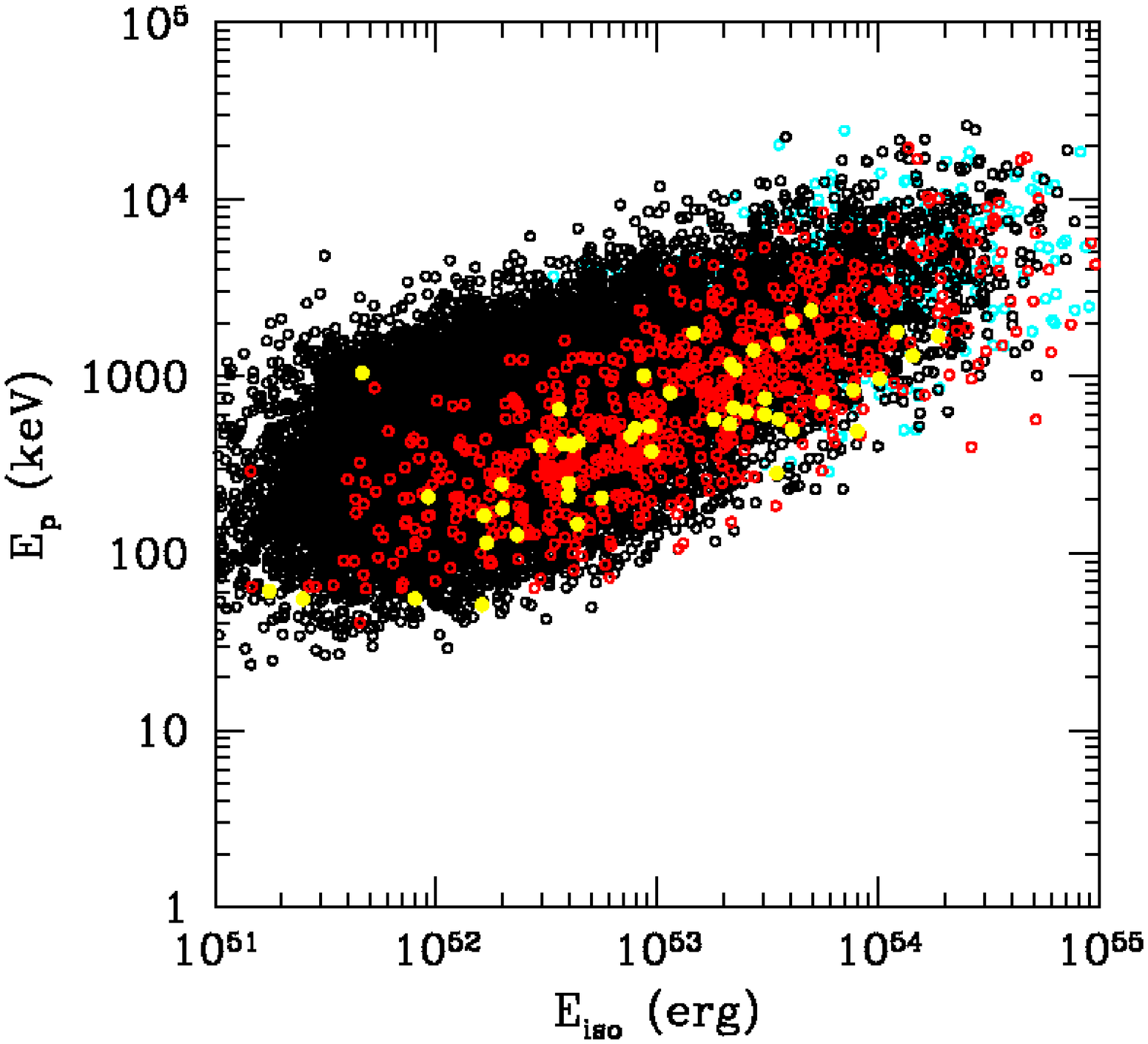} & 
\end{tabular}
\end{center}
\caption{$E_{\rm p}$ - $E_{\rm iso}$ relations with differents assumptions for the model parameters; black dots: whole synthetic 
population, except (in cyan) 
bursts that are optically thick as a result of photon-photon annihilation; red dots:
detected bursts assuming a threshold of 2.6 ph.cm$^{-2}$.s$^{-1}$ between 15 and 150 keV; yellow dots: observed BAT6 sample
(Salvaterra et al. 2012). Upper left panel: no correlation among
model parameters. Upper right panel: $40\, {\dot E_{52}}^{1/2}<{\bar \Gamma}<750$. Bottom left panel:
$t_{\rm v}={\rm min}\,[\tau,({\bar \Gamma}/200)^{-2}\ {\rm s}]$.
Bottom right panel: $\kappa=3\,{\bar \Gamma}_2^{0.5}$.  
   }
\end{figure*}

\section{Discussion and conclusion}
A satisfactory agreement with the observed $E_{\rm p}$ - $E_{\rm iso}$ relation can be achieved in several of the considered cases, 
but this is possible only if several strong constraints on the model parameters are satisfied:
\begin{itemize}
\item A large fraction of the dissipated energy has to be injected into a very small fraction of electrons. The value $\zeta=3\,10^{-3}$
(for $\epsilon_e=0.3$) we used should be considered as a lower limit however. Detailed hydro-calculations \citep{DM2000} 
show that Eqs.(7) and (8) both underestimate the density
and the dissipated energy by a factor 3 to 5. This means that the required value for $\zeta$ can probably be increased
by a factor of a few (possibly up to 10), but nevertheless remains very low.
\item The contrast $\kappa$ between the maximum and minimum Lorentz factor should be restricted to a narrow interval 
because otherwise the function $\varphi(\kappa)$ in Eq.(13) varies too much. This constraint can be somewhat relaxed if the fraction
$\zeta$ of accelerated electrons does not remain constant, but rises together with the dissipated energy per unit mass $e$. 
Assuming, for example, $\zeta \propto e$, Eq.(9) would be replaced by $E_{\rm p}\propto \Gamma_{\rm f}\,\rho^{1/2} e^{1/2}$
\citep{BDD09}. The dependence of the
peak energy on the contrast would be reduced, resulting in a larger allowed interval for $\kappa$.        
\item If the average Lorentz factor increases as ${\bar \Gamma}\propto {\dot E}^{1/2}$ (with no other connection among the model parameters),
the $E_{\rm p}$ - $E_{\rm iso}$ relation is lost. The peak energy is found to decrease with increasing $E_{\rm iso}$.
If ${\bar \Gamma}\propto {\dot E}^{1/2}$ is only a lower limit of $\bar \Gamma$ for a given ${\dot E}$, the $E_{\rm p}$ - $E_{\rm iso}$ relation
can be recovered.  
\item When the time scale or amplitude of the Lorentz factor variability is correlated with the average Lorentz factor (i.e.,
$t_{\rm v}\propto {\bar \Gamma}^{-2}$ or $\kappa\propto {\bar \Gamma}^{\nu}$ with $\nu\sim 0.5$, 
satisfactory $E_{\rm p}$ - $E_{\rm iso}$ relations are obtained. 
\item In all cases, the dispersion of the intrinsic $E_{\rm p}$ - $E_{\rm iso}$ correlation (i.e., excluding instrument threshold) is higher
than observed: the model accounts for the lack of bursts with a high $E_{\rm iso}$ and a low $E_{\rm p}$, but 
selection effects are responsible for the suppression of bursts with a low $E_{\rm iso}$ and a high $E_{\rm p}$. For case
({\it v}) the intrinscic $E_{\rm p}$ - $E_{\rm iso}$ relation is closest to the observed relation with 
$E_{\rm p}=169\,E_{\rm iso,52}^{0.57}$ keV and a dispersion of 0.39 in ${\rm Log}\,E_{\rm p}$.  

\end{itemize}
These aforementioned conditions concern both {\it (i)} the dynamics of the flow and {\it (ii)} the redistribution of the dissipated energy: 
\begin{itemize}
\item
{\it (i)} In the first model we assumed that the parameters controlling the dynamics of the flow, $\dot E$, $\bar \Gamma$, $\kappa$, $\tau$ and $t_{\rm v}$
were not correlated. This model provides a reasonable fit of the observed $E_{\rm p}$ - $E_{\rm iso}$ relation with a dispersion 
that might be too large, however. We then tested a few possible
correlations: $\bar \Gamma\propto {\dot E}^{1/2}$, $t_{\rm v}\propto {\bar \Gamma}^{-2}$, $\kappa\propto {\bar \Gamma}^{0.5}$, etc. Only the first correlation,
if applied alone, does not yield acceptable results. Of all the constraints on the dynamics, the limited interval of acceptable values for the contrast in Lorentz 
factor appears to be the most restrictive.
\item
{\it (ii)} The redistribution of the shock-dissipated energy should be efficient with $\epsilon_e=0.1\ -\ 0.3$, and concern a very small fraction 
$\zeta=10^{-3}\ -\ 10^{-2}$ of the electron population. This is probably the most severe constraint on the model, with the related question 
of the radiative contribution of the rest of the population with a quasi-thermal distribution. 
\end{itemize}

The purpose of this short paper was to compare the predictions of the internal shock model with the $E_{\rm p}$ - $E_{\rm iso}$ relation, 
that is, assuming that internal
shocks are responsible for the prompt emission of GRBs, are they able to account for the relation, and conversely, what are the constraints imposed on
the model if the $E_{\rm p}$ - $E_{\rm iso}$ relation applies. We obtained constraints on both the dynamics of the 
flow and the microphysics, some of them appearing quite stringent. In all cases, except possibly when the contrast in Lorentz factor increases with
average Lorentz factor of the flow, selection effects are required to exclude events with a low $E_{\rm p}$ and a high $E_{\rm iso}$ 
and reproduce the observed correlation.   

\begin{acknowledgements} 
LN was supported 
by a Marie Curie Intra-European Fellowship of the European
Community’s 7th Framework Programme (PIEF-GA-2013-627715), by an ERC advanced grant
(GRB) and by the I-CORE Program of the PBC and the ISF (grant 1829/12). We also thank the 
French Program for High Energy Astrophysics (PNHE) for financial support.
\end{acknowledgements}

\bibliographystyle{aa} 
\bibliography{amati}

\end{document}